\documentclass[sigconf]{acmart}

\AtBeginDocument{%
  }

\copyrightyear{2026}
\acmYear{2026}
\setcopyright{cc}
\setcctype{by-nc-nd}
\acmConference[WebSci '26]{18th ACM Web Science Conference}{May 26--29, 2026}{Braunschweig, Germany}
\acmBooktitle{18th ACM Web Science Conference (WebSci '26), May 26--29, 2026, Braunschweig, Germany}
\acmDOI{10.1145/3795766.3799750}
\acmISBN{979-8-4007-2504-3/2026/05}

\usepackage{multirow}

\usepackage{amssymb}
\usepackage{makecell}
\usepackage{graphicx}
\usepackage{booktabs}
\usepackage{amsmath} 
\usepackage{subcaption}
\usepackage{xurl}
\usepackage[export]{adjustbox}

\newcommand{\dimos}[1]{\textcolor{black}{#1}}

\begin{document}



\title{Causal Effects of Trigger Words in Social Media Discussions: A Large-Scale Case Study about UK Politics on Reddit}


\author{Dimosthenis Antypas$^*$}\thanks{* Equal contribution}
\orcid{0000-0002-6880-5142}
\affiliation{%
  \institution{Cardiff University}
  \city{Cardiff}
  \country{United Kingdom}}
\email{AntypasD@cardiff.ac.uk}

\author{Christian Arnold$^*$}
\orcid{0000-0002-7042-594X}
\affiliation{%
  \institution{University of Birmingham}
  \city{Birmingham}
  \country{United Kingdom}}
\email{c.arnold.2@bham.ac.uk}

\author{Nedjma Ousidhoum}
\orcid{0000-0003-3015-4759}
\affiliation{%
  \institution{Cardiff University}
  \city{Cardiff}
  \country{United Kingdom}}
\email{COusidhoumN@cardiff.ac.uk}

\author{Carla Perez-Almendros}
\orcid{0000-0001-9360-4011}
\affiliation{%
  \institution{Cardiff University}
  \city{Cardiff}
  \country{United Kingdom}}
\email{PerezAlmendrosC@cardiff.ac.uk}

\author{Jose Camacho-Collados}
\orcid{0000-0003-1618-7239}
\affiliation{%
  \institution{Cardiff University}
  \city{Cardiff}
  \country{United Kingdom}}
\email{CamachoColladosJ@cardiff.ac.uk}

\renewcommand{\shortauthors}{Antypas et al.}


\begin{abstract}
Political debates on social media often escalate quickly, leading to increased engagement as well as more emotional and polarised exchanges. Trigger points \citep{mau2023triggerpunkte} represent moments when individuals feel that their understanding of what is fair, normal, or appropriate in society is being questioned. Analysing Reddit discussions, we examine how trigger points shape online debates and assess their impact on engagement and affect. Our analysis is based on over 100 million comments from subreddits centred on a predefined set of terms identified as trigger words in UK politics. We find that mentions of these terms are associated with higher engagement and increased animosity, including more controversial, negative, angry, and hateful responses. These results position trigger words as a useful concept for modelling and analysing online polarisation.
\end{abstract}





\received{10 December 2025}
\received[revised]{19 February 2026}
\received[accepted]{19 February 2026}

%

\maketitle


\section{Introduction}
Online debates may suddenly escalate, prompting users to engage in emotional and confrontational deliberations. While this exact moment is challenging to model and predict, it is essential in online communication. It signals when discussions derail and users shift from debating to attacking one another. Our study introduces trigger words as a useful concept for understanding why and when social media users engage in such behaviour\textemdash which is central to accounting for hate speech and other abusive online interactions.

\begin{figure}
    \centering
    \includegraphics[clip, trim=0.7cm 0.6cm 0.4cm 0.7cm,,width=0.38\textwidth]
    {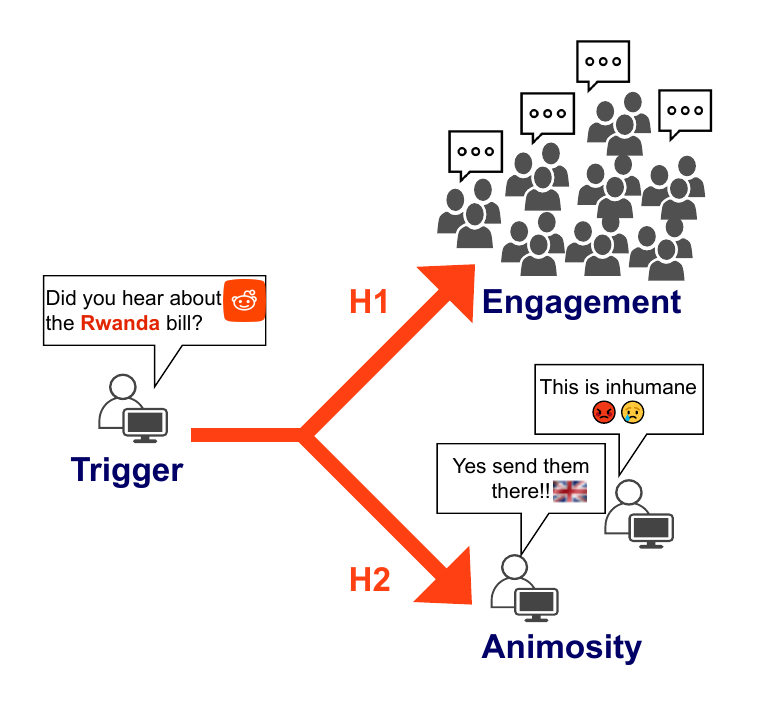}
    \caption{Consequences of trigger words. Hypothesis 1: Trigger words cause more engagement; Hypothesis 2: Trigger words cause more animosity.}
    \label{fig:triggerpoint_consequences}
\end{figure}

Trigger words are related to how individuals perceive themselves in society. People hold internal models of what counts as acceptable behaviour. These models include explicit, rule-like knowledge, e.g., formal norms or laws. More importantly, they also contain implicit, hard-to-articulate expectations about which actions are appropriate in a given context. Such deeply rooted beliefs about what is fair and what is unfair constitute a moral compass fundamental to navigating the social world. People feel triggered if these normative understandings are questioned (Figure~\ref{fig:triggerpoint_consequences}). More formally, \citet{mau2023triggerpunkte} define trigger points as ``\textit{those moments that question what is acceptable against the backdrop of individuals' understanding of the social contract. Trigger points jeopardise what individuals consider the fabric of society and their own role in it.}'' In our study, we investigate to what extent specific words in online discussions can behave as such trigger points. Trigger points can affect individuals through four mechanisms \citep{mau2023triggerpunkte}:

\noindent\textbf{Inequality.} When expectations about equality might not be met. Either similar people are treated unequally, or, to the contrary, people who should be treated differently are treated the same. For example, social benefits might be at eye level for minimum-wage earners. 

\noindent\textbf{Norm transgression.} It is the transgression of what is considered ``normal'' behaviour. A typical example relates to the lavish lifestyle of the privileged. 

\noindent\textbf{Slippery slope.} People may fear a society developing norms about appropriate behaviour that go in subjectively undesired directions. They may feel they lack the power to do anything about it. Claims such as ``Opening the borders so that a country could be flooded with immigrants'' are an example. 

\noindent\textbf{Behavioural demands.} A society may be perceived as demanding unreasonable behaviour. The role of pronouns is a case in point where the generic masculine pronouns ``he/him'' might trigger some, and the neutral pronoun ``they/them'' might trigger others.

Because these trigger points cause emotional discomfort, people respond in an affective-behavioural mode, increasing the debate's intensity and emotionality. In the social media context, this has two direct consequences (Figure~\ref{fig:triggerpoint_consequences}). Trigger words lead to: (i) \textbf{Hypothesis 1:} more user engagement, which leads to more messages or comments; and (ii) \textbf{Hypothesis 2:} higher levels of animosity, which causes polarisation, negativity, anger, and hate.

The goal of our study is to highlight and establish the concept of trigger words in scholarly work on online communication. Our research design and case selection closely follow this logic: We want to show that even the mention of a single trigger word can escalate online debates. To this end, we select five words that are \emph{ex-ante} likely to work as trigger words in the context of UK politics (\textit{Rwanda, Brexit, NHS, Vaccine, Feminism}). Collecting $>$100 million in related Reddit comments, we utilise text analytics and classifiers to determine whether these words elicit potentially negative, emotional, or harmful responses. The analysis builds on a systematic comparison between a treatment and control group in three different scenarios: (1) \textbf{Space:} specific subreddits where the selected words are likely to trigger users; (2) \textbf{Contextual:} words that appear in a similar context to the trigger words but without the triggering component; and (3) \textbf{\dimos{Temporal}:} periods identified as relevant for each specific trigger word. 


\section{Related Work}
Existing work studies how particular political words or phrases lead to more engagement \cite{Brady17, Hameleeres21} and posts with more emotion and animosity \cite{friedman2004positive, Martin_Huck_Williams_Hull_Hull_2021, Schroth2005stones, Rho20} in various online settings. However, they lack a joint conceptual and theoretical framework\textemdash something we aim to provide.
Trigger words are fundamental to understanding polarisation and division in online discussions. Given that they are at the heart of different conflictive behaviours later on, they help understand the causes of disagreements in online conversations \cite{de-kock-vlachos-2021-beg2} and their escalation \cite{de-kock-vlachos-2022-disagree,van-der-meer-etal-2023-differences} triggering explicit and implicit \textit{ad hominem} attacks, hate speech, abusive or offensive language on social media \cite{Talat2016HatefulSO, davidson2017automated,basile-etal-2019-semeval2,ousidhoum-etal-2019-multilingual2}.
Similarly, given a broad array of harmful online behaviours, the wide range of elements they involve, and how they can be included in social media data \cite{olteanu2019social}, \citet{blodgett-etal-2020-language2} report a missing paradigm that helps assess the initial causes without mainly focusing on the performance of the detection model. Studying trigger words helps track the origins of such behaviour, assess whether a social media post can lead to toxic comments \cite{dahiya2021would}, and how language affects polarisation on social media \cite{Karjus2024}. It lays the groundwork for understanding how polarised rhetoric may emerge and lead to intergroup conflict \cite{Almog22}.

Trigger words resemble dog whistles\textemdash coded expressions with dual meanings, one benign and one provocative to insiders \cite{breitholtz-cooper-2021-dogwhistles2}. Originating in US racial politics \cite{lopez2013dog}, the dog whistle concept covers systemic racism \cite{Filimon03032024}, political ambiguity in identity discourse \cite{Tolvanen24}, religion \cite{perry2023mating}, and immigration \cite{filimon2016dog}. Dog whistles are used deliberately to obscure intent. Trigger words, however, are not covert but evoke strong affective reactions and heighten perceived otherness.

\begin{table}[]
\resizebox{\columnwidth}{!}{
\begin{tabular}{lccccc}
\centering
                    & \textbf{Rwanda} & \textbf{Brexit} & \textbf{NHS} & \textbf{Vaccine} & \textbf{Feminism} \\ \hline
\textbf{Inequality}          &  \checkmark     &    \checkmark   &  \checkmark  &    \checkmark    &           \\ \midrule
\textbf{Norm transgression}  &        &    \checkmark   &     &         &    \checkmark      \\ \midrule
\textbf{Slippery Slope}      &  \checkmark     &    \checkmark   &     &    \checkmark    &    \checkmark      \\ \midrule
\textbf{Behavioural Demands} &        &        &     &    \checkmark    &    \checkmark      \\ \bottomrule

\end{tabular}}
\caption{Our trigger words and their respective trigger mechanisms (see introduction).}
\label{tab:trigger_word_choices}
\end{table}

Research on automating the identification of affective triggers in online discussions remains limited. For instance, \citet{almerekhi2022provoke} developed a framework to identify triggers for toxic behaviour within Reddit discussions. Similar work typically uses a restricted set of lexical features, primarily sentiment analysis, and often examines the level of engagement some topics may generate \cite{aldous2023really}. Others, such as \citet{beel2022linguistic}, explicitly investigate highly controversial topics, including abortion, climate change, and gun control, attempting to predict contentiousness within discussions. They often heavily rely on Reddit comment upvote ratios as a proxy for contentiousness. Finally, some studies focus on online conversational tensions by predicting conversational derailment, considering the overall progression of a conversation rather than specific terms as causes \cite{zhang2018conversations,hessel2019something}. 

\section{Methodology}
To empirically demonstrate that individual words \textbf{can} act as triggers in online debates, we look into the most likely cases in the context of UK politics. Our goal is to report behavioural mechanisms similar to those observed by \citet{mau2023triggerpunkte} offline, i.e., (1)\ more engagement and (2)\ animosity. If we could \textbf{not} find these effects in these intuitive cases, we would falsify the theory for online contexts.

\subsection{Term selection}\label{subsec:term_selection}
We selected \textit{National Health Service (NHS)}, \textit{Brexit}, \textit{Rwanda}, \textit{Vaccine}, and \textit{Feminist/Feminism}. These terms are associated with high-profile public debates, policy changes, and social movements in the UK. Hence, they typically provoke significant public engagement and emotional responses, i.e., the intended reactions in British social media users. Table~\ref{tab:trigger_word_choices} suggests mechanisms through which the terms may trigger individuals.

\noindent\textbf{Rwanda.}
Until late 2021, Rwanda was simply one of many African countries in the eyes of prevailing UK public discourse. However, in early 2022, the term's meaning became emblematic of how the British government suggested handling asylum seekers, and the home secretary struck a deal with Rwanda: the \textit{Migration and Economic Development Partnership}.\footnote{https://www.gov.uk/government/publications/uk-rwanda-treaty-provision-of-an-asylum-partnership/uk-rwanda-treaty-provision-of-an-asylum-partnership-accessible} With this agreement, \textit{Rwanda} became a trigger word for discussions on immigration policy in the UK from 2022 onwards, particularly related to ``unfair favours towards migrants and unregulated immigration'' \cite{economist2023rwandan}.

\noindent\textbf{Brexit.}
The UK's departure from the European Union has been a contentious issue since its inception. The division among the population was already evident during the tightly contested 2016 referendum \cite{ARNORSSON2018301}. Additionally, misinformation and disinformation have intensified these debates \cite{holler2021human,bruno2022brexit}. Considerations about treating EU migrants differently from the national population (inequality), norm transgressions of ``criminal'' migrants, or opening the gate to ``unlimited'' immigration make \textit{Brexit} a potential trigger word.

\noindent\textbf{NHS.}
Due to financial austerity measures, the UK government was accused of reducing its budget and neglecting the healthcare system \cite{guardian2022nhsbacklog, independent2024nhs}. Some support the claim that the NHS receives adequate funding, while others argue that it wastes resources or that its services are in decline \cite{gershlick2015public}. This trigger word is related to going against considerations about unequal treatment.

\noindent\textbf{Vaccine.} Recently, public vaccination programs, particularly those against COVID-19, have caused significant debate. Post-roll-out, debates have shifted to focus on the fairness of vaccine distribution and unreasonable demands for vaccination \cite{nichol2021ethics}. Moreover, the proliferation of misinformation and conspiracy theories has fueled public division \cite{hayawi2022anti, garett2021online}, particularly in social media. 

\noindent\textbf{Feminism.} 
The anonymity provided by social media often emboldens users to express misogyny \cite{barker2019online}. Some influencers have built a large audience by sharing and promoting harmful content against women. This has led to online echo chambers \cite{farrell2019exploring} where such views are amplified and propagated. We include \textit{Feminism} as a trigger word based on mechanisms related to norm transgressions, fears of a slippery slope, and unreasonable behavioural demands. 

\begin{table}[]
\resizebox{\columnwidth}{!}{
\begin{tabular}{ll}
\toprule
\textbf{Keyword} & \multicolumn{1}{c}{\textbf{Subreddits}} \\ \hline
\textbf{Feminism} & \begin{tabular}[c]{@{}l@{}}AskMen, MensRights, \\ antifeminists, AndrewTateTalk\end{tabular} \\ \hline
\textbf{Brexit} & \begin{tabular}[c]{@{}l@{}}neoliberal, brexit, ireland, unitedkingdom, \\ The\_Donald, ukpolitics, tories, \end{tabular} \\ \hline
\textbf{NHS} & \begin{tabular}[c]{@{}l@{}}neoliberal, CoronavirusUK, unitedkingdom,\\ ukpolitics, tories, labourUK\end{tabular} \\ \hline
\textbf{Rwanda} & \begin{tabular}[c]{@{}l@{}} neoliberal, tories, labourUK, currenteventsuk\\ ukpolitics, uknews, unitedkingdom \end{tabular} \\ \hline
\textbf{Vaccine} & \begin{tabular}[c]{@{}l@{}}conspiracy, neoliberal, unpopularopinion, \\ antifascistsofreddit, coronavirusuncensored, \\  CoronavirusUK, Coronavirus, wuhan\_flu  \end{tabular} \\ \hline
\end{tabular}}
\caption{Subreddits for each trigger word. }
\label{tab:keywords_subreddits}
\end{table}

\begin{table*}[ht]
\centering
\resizebox{\textwidth}{!}{
\begin{tabular}{lcccc|cccc|cccc|cccc|ccrr}
\toprule
 & \multicolumn{4}{c|}{\textbf{Feminist}} & \multicolumn{4}{c|}{\textbf{Rwanda}} & \multicolumn{4}{c|}{\textbf{Brexit}} & \multicolumn{4}{c|}{\textbf{NHS}} & \multicolumn{4}{c}{\textbf{Vaccine}} \\
 & \textbf{Be} & \textbf{Af} & \textbf{DiD} & \textbf{CI} & \textbf{Be} & \textbf{Af} & \textbf{DiD} & \textbf{CI} & \textbf{Be} & \textbf{Af} & \textbf{DiD} & \textbf{CI} & \textbf{Be} & \textbf{Af} & \textbf{DiD} & \textbf{CI} & \textbf{Be} & \textbf{Af} & \multicolumn{1}{c}{\textbf{DiD}} & \multicolumn{1}{c}{\textbf{CI}} \\ \hline
\textbf{\begin{tabular}[l]{@{}l@{}}Treatment\\ Only\end{tabular} } & 47 & 53 & \multicolumn{2}{c|}{} & 45 & 55 & \multicolumn{2}{c|}{} & 46 & 54 & \multicolumn{2}{c|}{} & 46 & 54 & \multicolumn{2}{c|}{} & 47 & 53 & \multicolumn{1}{c}{} & \multicolumn{1}{c}{} \\ \hline
\textbf{Space} & 49 & 51 & \textbf{4.1} & (3.8, 4.5) & 50 & 50 & \textbf{9.7} & (8.2, 11.3) & 49 & 51 & \textbf{6.3} & (6.1, 6.6) & 49 & 51 & \textbf{6.1} & (5.7, 6.4) & 49 & 51 & \textbf{4.1} & (3.5, 3.9) \\
\textbf{Contextual} & 48 & 52 & \textbf{4.6} & (3.9, 5.2) & 50 & 50 & \textbf{9.8} & (6.6, 14.9) & 47 & 53 & \textbf{3.2} & (2.6, 3.7) & 53 & 47 & \textbf{13.4} & (8.7, 18.1) & 49 & 51 & \textbf{5.0} & (3.1, 6.9) \\
\textbf{Temporal} & \multicolumn{4}{c|}{-} & 49 & 51 & \textbf{7.2} & (2.9, 11.6) & \multicolumn{4}{c|}{-} & \multicolumn{4}{c|}{-} & 50 & 50 & \textbf{5.9} & (4.1, 7.6) \\ \bottomrule
\end{tabular}}
\caption{Percentage of comments per thread before (Be) and after (Af) each trigger appearance (Hypothesis 1). Difference-in-Difference (DiD) and 95\% confidence interval (CI) included for the Control, Space, and Contextual comparisons. Bold scores indicate statistically significant results.} 
\label{tab:counts}
\end{table*}

\subsection{Data collection}
\label{sec:data_collection}
We collected English-language Reddit comments via the Pushshift API \cite{baumgartner2020pushshift}. Searching for exact matches of specific keywords in comments from 2019 to 2022, we collected the entire discussion thread for each extracted comment.\footnote{When extracting comments for the term \textit{woman}, we capped the maximum number of comments at 20 million due to the large volume.} 
We excluded comments in languages other than English using a fastText-based language identifier \cite{bojanowski-etal-2017-enriching2} and also removed comments that the API marked as ``[deleted]''. Only threads with comments both before and after the target keywords were retained to enable analysis of pre- and post-trigger differences.

As a final step, we included only comments posted within 30 minutes before and after the first mention of the trigger word. This window captures the most relevant responses, given that Reddit interactions peak quickly and early replies often shape the conversation’s direction \cite{thukral2018analyzing,singer2016evidence,horne2017identifying}.

\subsection{Research design}
We estimate the causal effect of trigger words using the difference-in-differences (DiD) estimator, which systematically compares a treatment group with a control group \citep{angrist2009mostly, Card1994}. It calculates the causal effect of an intervention as the difference in the treatment group before and after the intervention minus the difference in the control group before and after the intervention. In our context, we expect users in the treatment group to be highly likely to be triggered by the respective terms, whereas those in the control group are expected to be much less likely to be triggered by the exact words. 

\subsubsection{Control groups}
Discussions on Reddit are organised in different fora (subreddits). We systematically compare the treatment and control groups across three dimensions.\footnote{Identifying the causal effect in three completely unrelated comparisons helps address concerns about eventual endogeneity.}

\noindent\textbf{1. Space:}
To investigate the effects of trigger words on specific subsets of users, we define the treatment group as a set of popular subreddits related to the trigger word (Table \ref{tab:keywords_subreddits}). We compare them with discussions in a control group, i.e., the rest of Reddit, which includes many non-domain-specific subreddits. 

\noindent\textbf{2. Contextual:} 
We also compare trigger words to similar terms that lack the deeper transgressive meaning. For users in the subreddits listed in Table~\ref{tab:keywords_subreddits}, the difference between a trigger word and a milder control term should make a difference. We define control terms by pairing, for example, woman with feminism/feminist, Red Cross with NHS, EU/European Union with Brexit, surgery with vaccine, and Rwanda with its neighbouring countries\footnote{Burundi, Tanzania, Congo, Democratic Republic of Congo, Uganda, Kenya, and Zambia.}. Treatment and control data come from the same subreddits and year.

\noindent\textbf{3. Temporal:}
In the third experiment, we use the fact that two of our terms\textemdash \textit{Vaccine} and \textit{Rwanda}\textemdash only became explicit trigger words at specific points in time. For ``Rwanda'', we compare discussions in subreddits likely to spark reactions in 2020 with those in 2022. For \textit{Vaccine}, we compare threads from 2019 (pre-COVID-19 pandemic) to 2022 (post-COVID-19 pandemic). Again, we expect that users from the subreddits listed in Table~\ref{tab:keywords_subreddits} were triggered only in 2022, not in earlier years. The treatment and control groups share the same subreddits but differ in year.

\subsubsection{Statistical analysis}
We estimate the trigger words' treatment effect in 60 separate experiments\textemdash 12 experiments for H1 (Table~\ref{tab:counts}) and 48 experiments for H2 (Table~\ref{tab:results}). For each experiment, we prepare the data in the same way. The treatment is the first mention of a trigger word in a discussion thread, i.e., there is only one well-defined treatment per thread. To test H1, we count the number of comments for each discussion thread 30 minutes before and 30 minutes after the treatment in the treatment group and the placebo treatment in the control group. We then calculate the share of comments before and after the intervention. For example, if there are 8 comments during the 30 minutes before the trigger word and 12 comments during the 30 minutes after the trigger word, we normalise the data of this discussion thread to 40\% before intervention and 60\% after intervention. For experiments with H2, we count the number of comments expressing animosity in a thread and then normalise the data as a share of animosity per thread before and after the trigger word.

We identify the causal DiD-estimator with a two-way fixed effects model in an OLS-regression framework \cite{angrist2009mostly}: \textit{$\text{proportion} \sim \text{trigger} + \text{treatment} + \text{interaction}$}
where \textit{proportion} is the per-thread-normalised target feature, and the interaction term is given by
\textit{$\text{interaction} = \text{trigger} \times \text{treatment}.$}
Since the DiD estimator subtracts the difference before and after intervention in the control group from the equivalent in the treatment group, no further control variables are necessary \textemdash if the parallel trends assumption holds \cite{roth2023s}. In the absence of treatment, the difference in outcomes between the treated and control groups would follow the same trajectory over time.\footnote{See Section~\ref{sec:robustness_main} for further explanations and Appendix \ref{sec:parallel_trends} for full detailed analyses.} 

\subsection{Features} \label{sec:features}
We rely on comment counts to test an increase in user engagement (H1). To measure animosity (H2), we identify relevant features using various language models based on the RoBERTa architecture \cite{liu2019roberta} that are specialised for social media. In Appendix \ref{sec:appendix-model_selection}, we validate the proposed approach through a manual evaluation on a random Reddit sample across a variety of open-source transformer-based language models.

\subsubsection{User engagement (Hypothesis 1)}
Our first hypothesis posits that trigger words increase user engagement. We rely on the number of comments to investigate this and expect an uptick in message frequency after the trigger words are mentioned for the first time in a thread.

\subsubsection{Animosity (Hypothesis 2)}
Hypothesis 2 suggests that trigger words cause more animosity in a debate. We operationalise this with controversial comments, negative sentiment, anger, and different forms of hate speech.

\noindent\textbf{Controversiality.} Reddit itself identifies a comment as controversial if it receives similar amounts of upvotes and downvotes. We expect trigger words to cause more controversy. 

\begin{figure}
    \centering
    \resizebox{\columnwidth}{!}{
    \includegraphics{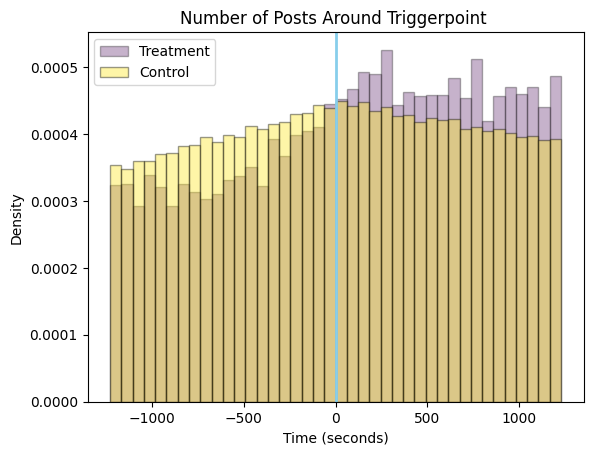}
    }
    \caption{Example: Distribution of comments in the treatment and control groups (space) for the trigger word \textit{Rwanda}.}
    \label{fig:counts_space_rwanda}
\end{figure}

\noindent\textbf{Sentiment analysis.} We use \textit{twitter-roberta-base-sentiment-latest} \cite{loureiro-etal-2022-timelms2} to identify a comment as \textit{negative, neutral, or positive}. This model has been fine-tuned for sentiment analysis using the SemEval 2017 Sentiment Analysis dataset \cite{rosenthal2019semeval}. Our approach builds on existing work that uses sentiment analysis to study online debates on controversial topics such as vaccination \cite{melton2021public}, politics \cite{guimaraes2019analyzing}, and gender bias \cite{marjanovic2022quantifying}.

\noindent\textbf{Negative emotions.} 
Using \textit{twitter-roberta-base-emotion-multilabel-latest} \cite{camacho2022tweetnlp}, we assign one or more emotions to each tweet. This model is trained on the Semeval 2018 Affect in Tweets task \cite{mohammad2018semeval} and covers 11 different emotions. In our analysis, we expect to observe increased anger following a trigger word.

\noindent\textbf{Hate speech.} Finally, we use the \textit{twitter-roberta-base-hate-multi-class} model \cite{antypas2023robust} to detect hate speech. It is trained on a combination of 13 hate-speech Twitter datasets and predicts hate speech directed at seven target groups. Hate speech detection is motivated by previous research revealing hateful trends in different topics \cite{farrell2019exploring, rieger2021assessing, walter2022vaccine}. For \textit{NHS}, \textit{Brexit} and \textit{Vaccine}, we consider any hateful comments. We focus on sexism for \textit{Feminism/Feminist}, and on racism for \textit{Rwanda}.

\section{Results}
\subsection{User engagement (Hypothesis 1)}
According to our first hypothesis, trigger words cause more comments. The DiD estimate reveals significant differences in comment frequency between the treatment and control groups for all our trigger words (Table~\ref{tab:counts}).  

In the treatment-only subset, the difference in user engagement before and after the trigger word is evident across all terms. It is the largest, with \textit{Rwanda}, where we observe a 10\%-point increase (Table~\ref{tab:counts}). To identify the causal effect of the trigger word, we compare the difference in the treatment group to the difference in the control group (DiD estimate). Across all settings, we observe a significant difference in the number of comments between the treatment and control subsets, with the treatment group showing a larger increase. The lowest value is a 3.2 percentage-point increase in the contextual comparison for \textit{Brexit}. The largest increase is for \textit{NHS} in the contextual control group, with a 13.4\% point increase.

When examining comment frequency across various control groups, the impact of specific trigger words becomes more apparent. For instance, Figure \ref{fig:counts_space_rwanda} illustrates the distribution of comments related to the trigger \textit{Rwanda} in the space-related experiment with the overall number of messages across all threads. This trigger word leads to varying levels of engagement across different Reddit communities, with a notable increase in comments in identified communities where Rwanda is more controversial. Similar patterns are also observed in the contextual and temporal control groups for \textit{Rwanda} (Appendix~\ref{sec:appendix_counts}, Figure~\ref{fig:rwanda_temporal_semantic}). Overall, our analysis of the influence of trigger words across different control groups indicates a clear increase in engagement for each trigger word analysed, confirming our first hypothesis.

\begin{table*}[t!]
\begin{minipage}{0.45\textwidth}
\centering
\resizebox{0.98\textwidth}{!}{
\begin{tabular}{cc|ccc}
\textbf{Trigger} & \textbf{Setting} & \textbf{Feature} & \textbf{DiD} & \textbf{CI} \\ \hline
\multirow{8}{*}{\rotatebox{90}{Brexit}} & \multirow{4}{*}{\rotatebox{90}{Space}} & Contr & \textbf{1.0} & (0.8, 1.3) \\
 &  & N.Sent & \textbf{3.3} & (3.0, 3.5) \\
 &  & Anger & \textbf{2.7} & (2.4, 3.0) \\
 &  & Hate & -0.1 & (-0.5, 0.4) \\ \cline{2-5} 
 & \multirow{4}{*}{\rotatebox{90}{Context.}} & Contr & \textbf{1.4} & (0.7, 2.1) \\
 &  & N.Sent & \textbf{2.6} & (2.0, 3.1) \\
 &  & Anger & \textbf{3.8} & (2.1, 5.5) \\
 &  & Hate & -0.5 & (-1.9, 0.9) \\ \hline
\multirow{8}{*}{\rotatebox{90}{Feminist}} & \multirow{4}{*}{\rotatebox{90}{Space}} & Contr & 0.3 & (-0.2, 0.8) \\
 &  & N.Sent & \textbf{2.6} & (2.2, 3.0) \\
 &  & Anger & \textbf{1.4} & (1.0, 1.9) \\
 &  & Sexism & \textbf{0.6} & (0.1, 1.1) \\ \cline{2-5} 
 & \multirow{4}{*}{\rotatebox{90}{Context.}} & Contr & 0.1 & (-0.8, 1.1) \\
 &  & N.Sent & \textbf{3.0} & (2.3, 3.6) \\
 &  & Anger & \textbf{1.4} & (1.0, 1.9) \\
 &  & Sexism & \textbf{1.4} & (0.5, 2.3) \\ \hline
\multirow{8}{*}{\rotatebox{90}{NHS}} & \multirow{4}{*}{\rotatebox{90}{Space}} & Contr & \textbf{1.1} & (0.7, 1.4) \\
 &  & N.Sent & \textbf{3.1} & (2.8, 3.5) \\
 &  & Anger & \textbf{2.7} & (2.3, 3.1) \\
 &  & Hate & -0.3 & (-0.9, 0.3) \\ \cline{2-5} 
 & \multirow{4}{*}{\rotatebox{90}{Context.}} & Contr & 4.5 & (-3.2, 12.2) \\
 &  & N.Sent & 3.5 & (-1.2, 8.3) \\
 &  & Anger & 4.2 & (-11.6, 20.0) \\
 &  & Hate & -1.4 & (-12.5, 9.6)
\end{tabular}
}
\end{minipage}
\hspace{0.05\linewidth}
\begin{minipage}{0.45\textwidth}
\centering
\resizebox{0.98\textwidth}{!}{
\begin{tabular}{cc|ccc}
\textbf{Trigger} & \textbf{Setting} & \textbf{Feature} & \textbf{DiD} & \textbf{CI} \\ \hline
\multirow{12}{*}{\rotatebox{90}{Rwanda}} & \multirow{4}{*}{\rotatebox{90}{Space}} & Contr & \textbf{2.3} & (0.8, 3.7) \\
 &  & N.Sent & \textbf{5.5} & (4.0, 6.9) \\
 &  & Anger & \textbf{4.0} & (2.7, 5.3) \\
 &  & Racism & \textbf{1.8} & (0.0, 3.6) \\ \cline{2-5} 
 & \multirow{4}{*}{\rotatebox{90}{Context.}} & Contr & \textbf{6.6} & (0.2, 13.0) \\
 &  & N.Sent & \textbf{5.5} & (1.5, 9.6) \\
 &  & Anger & 5 & (-4.6, 14.6) \\
 &  & Racism & 14.6 & (-0.7, 29.9) \\ \cline{2-5} 
 & \multirow{4}{*}{\rotatebox{90}{Temporal}} & Contr & \textbf{5.9} & (0.5, 11.3) \\
 &  & N.Sent & \textbf{4.8} & (0.8, 8.8) \\
 &  & Anger & 2.8 & (-1.0, 6.6) \\
 &  & Racism & 0.9 & (-11.1, 12.8) \\ \cline{1-5} 
\multirow{12}{*}{\rotatebox{90}{Vaccine}} & \multirow{4}{*}{\rotatebox{90}{Space}} & Contr & \textbf{1.5} & (1.3, 1.8) \\
 &  & N.Sent & \textbf{3.0} & (2.8, 3.2) \\
 &  & Anger & \textbf{2.0} & (1.7, 2.2) \\
 &  & Hate & \textbf{0.4} & (0.0, 0.8) \\ \cline{2-5} 
 & \multirow{4}{*}{\rotatebox{90}{Context.}} & Contr & \textbf{2.2} & (1.1, 3.4) \\
 &  & N.Sent & \textbf{5.4} & (4.7, 6.1) \\
 &  & Anger & \textbf{4.4} & (1.6, 7.2) \\
 &  & Hate & -0.7 & (-2.4, 0.9) \\ \cline{2-5} 
 & \multirow{4}{*}{\rotatebox{90}{Temporal}} & Contr & -0.3 & (-3.0, 2.3) \\
 &  & N.Sent & \textbf{2.9} & (1.4, 4.4) \\
 &  & Anger & \textbf{5.5} & (1.4, 9.6) \\
 &  & Hate & -2.0 & (-5.3, 1.3)
\end{tabular}
}
\end{minipage}
\caption{Difference in Difference (DiD) scores along with the 95\% confidence intervals (CI) for each setting tested. Bold scores indicate statistically significant results.}
\label{tab:results}
\end{table*}

\subsection{Animosity (Hypothesis 2)}
Our second hypothesis suggests that trigger words may increase the frequency of polarised messages and lead to more animosity in the discussion.\footnote{Examples of discussions are presented in Appendix \ref{sec:appendix-example}.}

\begin{figure}
    \centering
    \resizebox{\columnwidth}{!}{
    \includegraphics{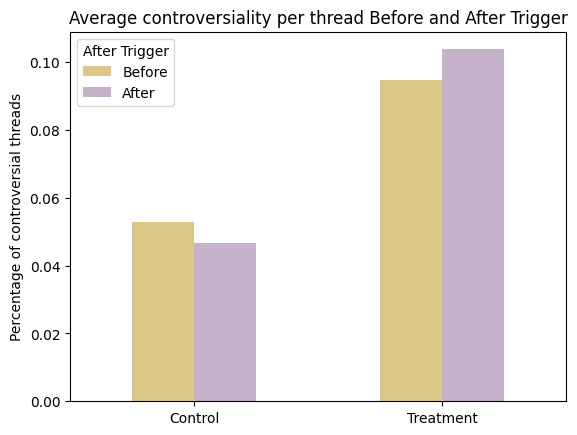}
    }
    \caption{Average controversiality per thread for the treatment and control (space) groups of \textit{Rwanda}.}
    \label{fig:controversiality_rwanda}
\end{figure}

\noindent\textbf{Controversiality.} We analyse the average controversiality of each thread and compare the proportion of controversial comments before and after the introduction of the trigger words. Our results (Table~\ref{tab:results}) indicate that, in general, there are more controversial comments after a trigger word. The largest difference is the contextual comparison for \textit{Rwanda}\textemdash a causal effect of 6.6\% points. However, not all differences are statistically significant. Notably, there is no increase in controversial comments for \textit{Feminism}, and in the temporal and contextual settings for \textit{Vaccine} and \textit{NHS}, respectively.

As an example, Figure~\ref{fig:controversiality_rwanda} shows the percentages of controversial threads related to \textit{Rwanda} for the space experiment, before and after its occurrence. The data indicates a noticeable increase in thread controversiality caused by the trigger word (DiD $= 2.3\%$ points). 

\noindent\textbf{Sentiment \& emotion.}
Based on our hypothesis, trigger words tend to evoke more negative emotions in users. In our analysis, we focus on negative sentiment, in particular anger.  

Table~\ref{tab:results} indicates a clear trend for the negatively charged comments of the treatment and control groups for each trigger word. We observe a significant increase in comments conveying a negative sentiment after the trigger word appears in 11 of the 12 settings tested, with the sole exception being the contextual comparison for \textit{NHS}. The largest effect in sentiment (DiD $= 5.5\%$ points) appears for the \textit{Rwanda} trigger word in the space and contextual experiments.


Trigger words also cause more anger. We observe positive causal effects across all conditions, with only 3 non-significant effects. Substantively, we measure a causal effect that ranges from 1.4\% points more anger (\textit{Feminist}) to 5.5\% points (\textit{Vaccine contextual comparison}).

\noindent\textbf{Hate speech.}
As a final indicator of animosity, we also consider hate speech and abusive language, more specifically any kind of toxicity for \textit{Brexit}, \textit{NHS}, and \textit{Vaccine}, and the more domain-specific racism-related hate speech for \textit{Rwanda} and sexism-related hate speech for \textit{Feminism}. As seen from Table \ref{tab:results}, the impact of trigger words regarding the occurrence of hateful or offensive comments is not as prevalent as anger and negative sentiment. We observe significant effects in areas where we can rely on a domain-specific hate speech measure, i.e., in \textit{Feminism} (space and contextual comparison) and \textit{Rwanda} (space group).



\section{Robustness of Results}
\label{sec:robustness_main}
We assess the robustness of our Difference-in-Differences (DiD) estimates through parallel trends tests and placebo treatments. The parallel trends assumption requires that, absent treatment, the treatment and control groups would follow similar trends over time. To assess this, we implement event studies that trace changes in the outcome variable across six 10-minute intervals before and after the first mention of a trigger word. As shown in Appendix~\ref{sec:parallel_trends}, Figure~\ref{fig:robustness}, pre-treatment trends remain flat and centred around zero, supporting the validity of the DiD design.

To further validate our identification strategy, we conduct placebo analyses by shifting the assumed treatment point in 10-minute increments before and after the actual intervention. For each placebo word, we re-estimate the DiD models. If our findings were artefacts of general conversational trends, we would expect similar effects with these alternative words. Instead, Figure~\ref{fig:placebos} in Appendix~\ref{sec:parallel_trends} shows that only the actual treatment time yields a pronounced effect, while estimates at placebo times converge toward zero. Together, our checks strengthen confidence in our causal claims.

\section{Conclusion}
Our work represents the first systematic computational study of trigger words in online debates. The research design and case selection are geared toward identifying such behaviour in the most likely cases, namely a large-scale analysis of UK politics on Reddit using five trigger words (\textit{Rwanda, Feminism, Brexit, NHS, and Vaccine}).

The results show that trigger words lead to more user engagement and higher levels of animosity. This confirms the first hypothesis (H1), which predicts increased user interaction upon encountering a trigger word. Regarding our second hypothesis (H2), which concerns the impact of trigger words on the animosity in ensuing discussions, we examine four features that may indicate such effects. Trigger words cause more controversiality, more negative sentiment, and more anger. Our findings also confirm that trigger words cause more hate speech.

From this work, we can certify that trigger words exist in online communication. Future work will analyse this phenomenon more comprehensively. We aim to broaden the scope, study the prevalence of trigger words across countries and languages, and better understand their behavioural consequences. We also strive to automate trigger-word detection to identify relevant words in online conversations when we are not familiar with them \emph{a priori}.

\section{Limitations}
We refrain from using large language models (LLMs) for feature extraction primarily because of the size of our dataset ($>$100 million entries). Using LLMs would have significantly increased the computational and financial cost and the ecological impact of our experiments. Moreover, the existing literature indicates that LLMs' zero-shot capabilities do not consistently outperform smaller models on specific social media tasks such as ours \cite{antypas-etal-2023-supertweeteval}.

Our study does not aim to map the prevalence of trigger words across online debates. The selected subreddits may not fully represent the broader range of discussions these trigger words could elicit across the Reddit community. Similarly, we acknowledge the limitations and potential biases of our work, as we investigate only a small set of trigger words in the UK and limit our data to English-language comments from a limited selection of subreddits. However, as the first study to analyse the validity of trigger words in online discussions, we are confident that our findings warrant further investigation. 

\section{Ethics Statement}
All data used in our research is sourced from publicly available information or collected using the official Reddit API. We present aggregated data without reporting sensitive information from individual users.

The automatic methods used can be abused by those with the power to repress people with opposing opinions. Hence, we provide a comprehensive study but do not share our data publicly to prevent potential misuse or commercial exploitation. Researchers interested in replicating our experiments are encouraged to contact us directly.

\bibliographystyle{ACM-Reference-Format}
\bibliography{anthology,custom}

\appendix
\section{Model Selection}
\label{sec:appendix-model_selection}
\subsection{Validation of Annotation Model}

We use computationally efficient models that can be deployed locally to extract signals relevant to our hypotheses, and then validate the annotation procedure. We sampled 100 Reddit comments from our dataset, 20 for each trigger word, and then hand-annotated sentiment, emotion, and the presence of hate speech with three experienced NLP and CSS researchers. 
Each comment was assigned a label by majority vote, and disagreements were resolved through discussion (see Table \ref{tab:annotated_data}). Given the difficulty of the annotation task, the annotations above show above-average agreement scores for sentiment and hate speech, with Fleiss' $\kappa$ scores of 0.52 and 0.56, respectively. For the multi-label emotion annotation task, where annotators had to choose from a set of 12 labels, the Krippendorff’s alpha score \cite{krippendorff2011computing} reached 0.24, which is in line with other similar multi-label tasks with a high number of labels \cite{mohammad2018semeval,antypas-etal-2022-twitter2,ousidhoum-etal-2019-multilingual2}.

\begin{table}[h]
\begin{tabular}{llll|ll}
\multicolumn{2}{c|}{Emotion}            & \multicolumn{2}{c|}{Hate Speech}       & \multicolumn{2}{c}{Sentiment}         \\ \hline
no emotion & \multicolumn{1}{l|}{29} & not hate     & 88     & negative     & 56    \\
anger      & \multicolumn{1}{l|}{26} & sexism       & 7      & neutral      & 37    \\
disgust    & \multicolumn{1}{l|}{23} & racism       & 3      & positive     & 7     \\
pessimism    & \multicolumn{1}{l|}{19} & other                & 2               & \multicolumn{2}{l}{\multirow{6}{*}{}} \\
anticipation & \multicolumn{1}{l|}{15} & \multicolumn{2}{l|}{\multirow{5}{*}{}} & \multicolumn{2}{l}{}                  \\
sadness    & \multicolumn{1}{l|}{9}  & \multicolumn{2}{l|}{} & \multicolumn{2}{l}{} \\
surprise   & \multicolumn{1}{l|}{5}  & \multicolumn{2}{l|}{} & \multicolumn{2}{l}{} \\
optimism   & \multicolumn{1}{l|}{5}  & \multicolumn{2}{l|}{} & \multicolumn{2}{l}{} \\
joy        & \multicolumn{1}{l|}{2}  & \multicolumn{2}{l|}{} & \multicolumn{2}{l}{}
\end{tabular}
\caption{Distribution of comments regarding emotion, sentiment, and hate speech categories on the manually annotated dataset.}
\label{tab:annotated_data}
\end{table}

We fine-tune the base version of BERT-cased, XLNet, and RoBERTa for sentiment analysis, emotion recognition, and hate speech detection on the same datasets as the models we used for our experiments \cite{rosenthal2017semeval,mohammad2018semeval,antypas2023robust}.

We further compare our results against five publicly available pre-trained models on Hugging Face. Specifically, two sentiment classifiers trained on Reddit data\footnote{ reddit-dBERT: \url{https://huggingface.co/akshataupadhye/finetuning-sentiment-model-reddit-data}, reddit-XLNET: \url{https://huggingface.co/minh21/XLNet-Reddit-Sentiment-Analysis} }, hateBERT \cite{caselli-etal-2021-hatebert2} and one emotion classifier\footnote{BERT-emotion: https://huggingface.co/ayoubkirouane/BERT-Emotions-Classifier}. All models are evaluated in the aforementioned manually labelled dataset.

Table \ref{tab:models_evaluation} presents the average micro F1-scores achieved by the models across each task.\footnote{Micro F1-score provide a more realistic representation of the results, given the limited sample and heavily imbalanced data.} The Twitter-RoBERTa models achieved the highest scores on all tasks. These scores are matched by XLNet-base for sentiment analysis and XLNet for emotion detection, and the fine-tuned models RoBERTa and XLNet for hate speech detection. Overall, the tested models performed comparably well, except for two off-the-shelf sentiment classifiers that achieved scores below 50\%. 

\begin{table}[h]
\begin{tabular}{l|c|c|c}
\hline
\textbf{}    & \textbf{Sentiment} & \textbf{Emotion} & \textbf{Hate} \\ \hline
t-RoBERTa    & \textbf{0.76}      & \textbf{0.50}    & \textbf{0.88}          \\
RoBERTa    & 0.75               & 0.47             & \textbf{0.88}          \\
XLNet      & \textbf{0.76}      & 0.45             & \textbf{0.88}         \\
BERT       & 0.74               & 0.40             & 0.85          \\
reddit-dBERT & 0.45               & -                & -             \\
reddit-XLNet & 0.31               & -                & -             \\
BERT-emotion & -                  & 0.45             & -             \\
hateBERT     & -                  & -                & 0.80          \\ 
Random     &  0.30                  & 0.20                & 0.07         \\ \hline
\end{tabular}
\caption{Average micro-F1 scores of the models tested for the sentiment analysis, emotion, and hate speech detection tasks. \textit{Random} indicates a naive baseline that randomly selects labels.}
\label{tab:models_evaluation}
\end{table}

\subsection{Annotated Categories}
\begin{itemize}
    \item Emotion Categories: \textit{anger, anticipation, disgust, fear, joy, love, optimism, pessimism, sadness, surprise, trust}
    \item Hate Speech Categories: \textit{not\_hate, sexism, racism, religion, other, sexual\_orientation, disability}
\end{itemize}

\section{Extended Results}
\label{sec:appendix_counts}


Table \ref{tab:comment_count} presents the number of comments included in each experiment for both the treatment and control groups.



\begin{figure*}[ht]
    \centering
    \begin{subfigure}[b]{1\columnwidth}
        \centering
        \includegraphics[width=\textwidth]{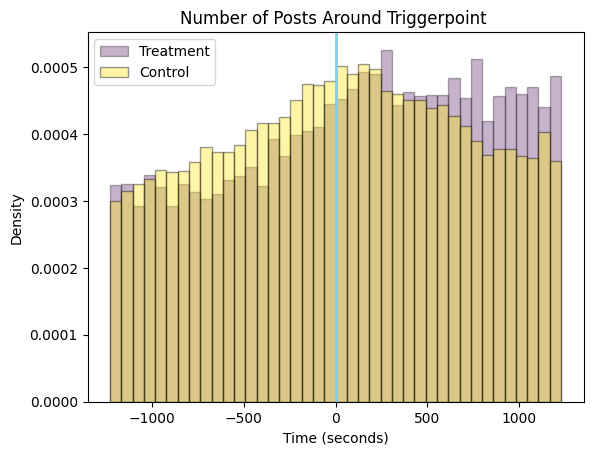}
        \caption{Contextual Comparison}
        \label{fig:image3a}
    \end{subfigure}
    \hfill
    \begin{subfigure}[b]{1\columnwidth}
        \centering
        \includegraphics[width=\textwidth]{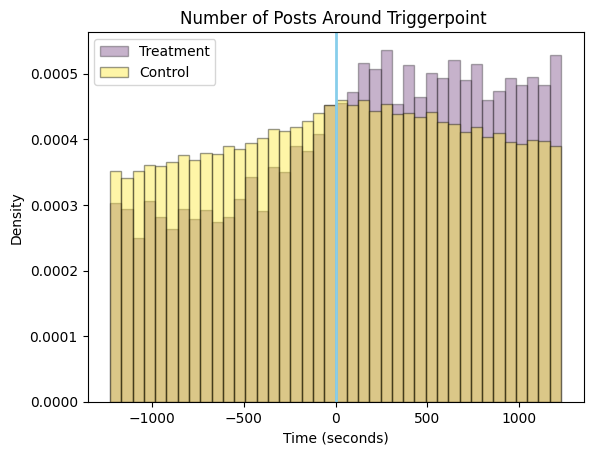}
        \caption{Temporal Comparison}
        \label{fig:image3b}
    \end{subfigure}

    \caption{Rwanda Treatment and Control distributions of comments in the Contextual and Temporal settings.}
    \label{fig:rwanda_temporal_semantic}
\end{figure*}

Figure \ref{fig:rwanda_temporal_semantic} displays the distribution of comments posted before and after the introduction of the \textit{Rwanda} trigger \dimos{word} in the Temporal and Contextual control settings.

\begin{table}[]
\centering
\begin{tabular}{l|lrr}
\multicolumn{1}{c|}{\textbf{Trigger}} & \multicolumn{1}{c}{\textbf{Experiment}} & \textbf{Treatment} & \textbf{Control} \\ \hline
\multirow{2}{*}{Feminist} & space    & \multirow{2}{*}{264,495} & 16,503,956 \\
                          & contextual &                         & 163,740   \\ \hline
\multirow{3}{*}{Rwanda}   & space    & \multirow{3}{*}{28,770}  & 668,641   \\
                          & contextual &                         & 2,587     \\
                          & temp     &                         & 5,896     \\ \hline
\multirow{2}{*}{Brexit}   & space    & \multirow{2}{*}{688,600} & 10,793,036 \\
                          & contextual &                         & 111,524   \\ \hline
\multirow{3}{*}{vaccine}  & space    & \multirow{3}{*}{348,641} & 7,685,095  \\
                          & contextual &                         & 12,986    \\
                          & temp     &                         & 11,841    \\ \hline
\multirow{2}{*}{NHS}      & space    & \multirow{2}{*}{509,458} & 5,275,917  \\
                          & contextual &                         & 1,276    
\end{tabular}
\caption{Number of comments present for each trigger in each experiment for both the treatment and control subsets.}
\label{tab:comment_count}
\end{table}

\section{Robustness of Results}
We conduct a series of robustness checks designed to test the sensitivity of our results to assumptions and alternative specifications.

\paragraph{Testing the Parallel Trends Assumption.}
\label{sec:parallel_trends}
The DiD estimator relies on the assumption of parallel trends. While there are no formal guarantees that the assumption holds, there are grounds to argue in its favour. The first argument is theoretical. In neither of our experimental designs did we have reason to believe that the frequency of expressing animosity would not follow the same trend without the trigger word. Whether comparing treatment and control groups across different Reddit subreddits (spaces), different words (contextual), or points in time (temporal), without the trigger word, the relative difference between treatment and control groups would remain the same.   

To further empirically support our argument, we conduct event studies. For each experiment, we model the outcome variable over time, enabling us to assess the dynamics of treatment impact before and after the intervention. Plotting the treatment effects at different time points before the treatment, the event study helps visualise whether the pre-treatment trends in the treatment and control groups are indeed parallel \cite{roth2023s}. 

For each of our 60 overall experiments, we divide the 30 minutes before and after the first mention of a trigger word into six 10-minute intervals, rather than just two. This allows us to track pre-treatment trends at higher granularity. We calculate each discussion thread's share of (hostile) messages in the six intervals. Averaging across all data points per period, we subtract the control group from the treatment group and compute the difference for each observation relative to the first observation period.

\begin{figure}
    \centering
    \resizebox{\columnwidth}{!}{
    \includegraphics{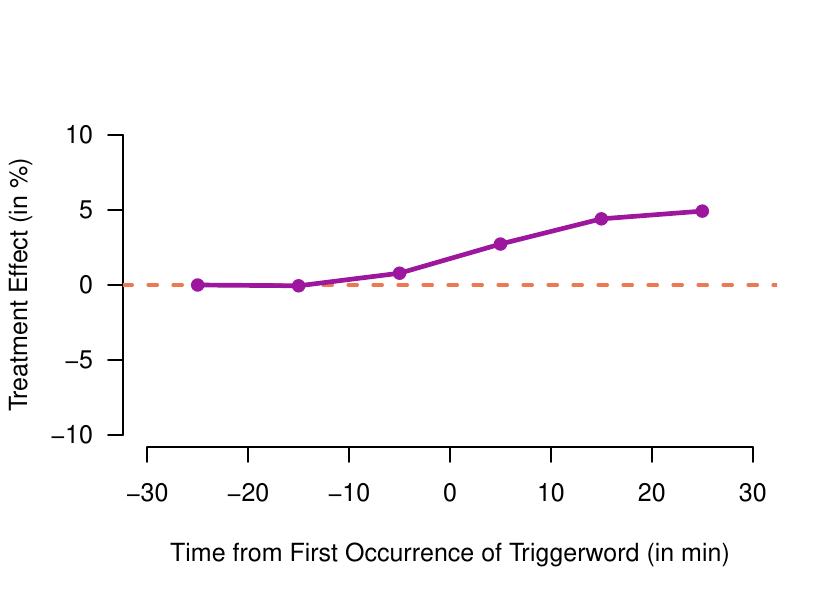}
    }
    \caption{Testing the Parallel Trends Assumption with event studies.}
    \label{fig:robustness}
\end{figure}

Figure~\ref{fig:robustness} charts the time-wise means over all 60 event studies. If the parallel trends assumption holds, the lines before the treatment at time $= 0$ should all remain on the horizontal line where the treatment effect $= 0$. Indeed, this is essentially what we find. The mean treatment effect is close to 0 before the trigger word is first mentioned. After the intervention, it is clearly above the horizontal 0-line. 


\begin{figure}
    \centering
    \resizebox{\columnwidth}{!}{
    \includegraphics{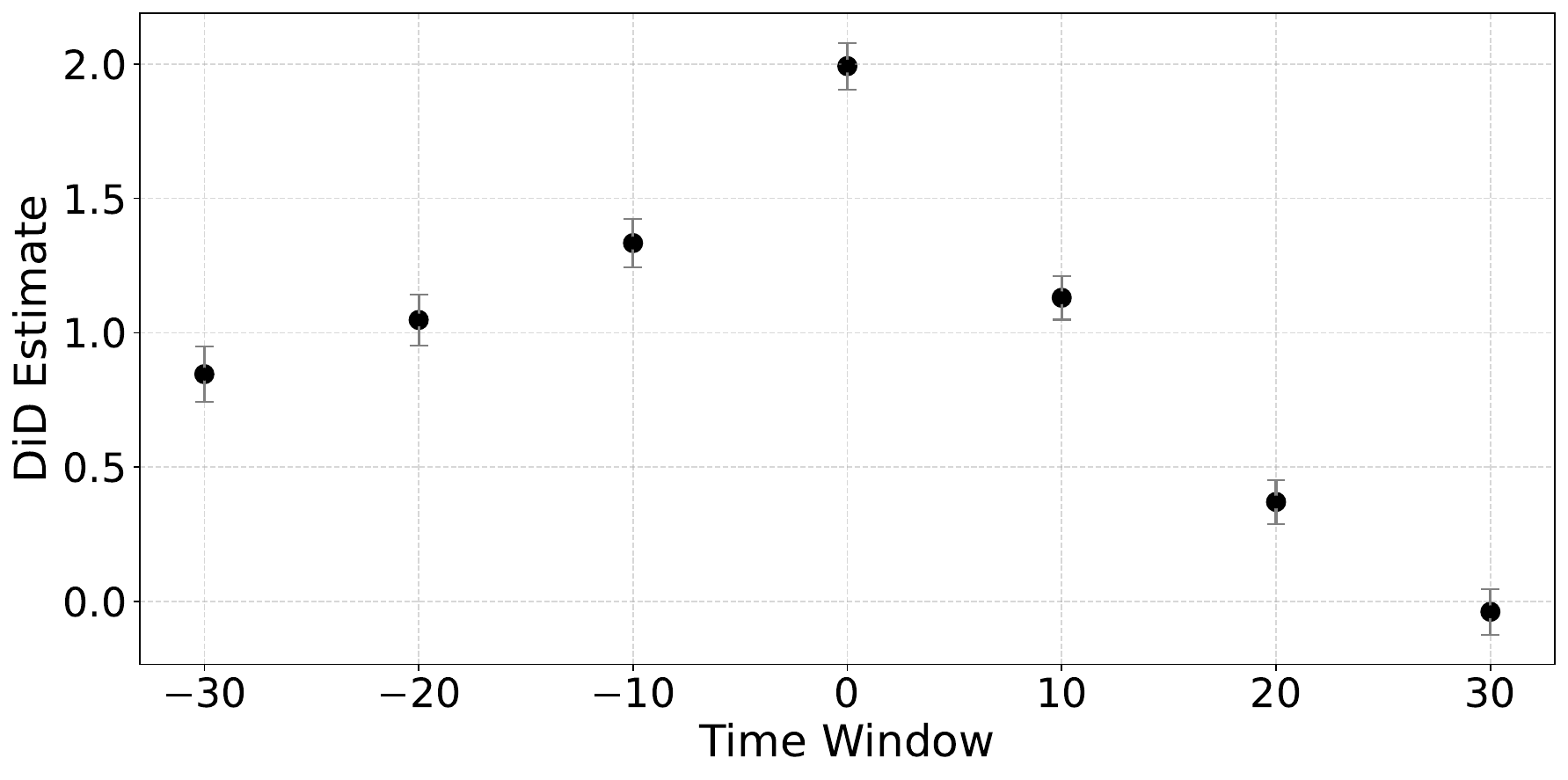}
    }
    \caption{Placebo treatments every 10 minutes, ranging from 30 minutes before to 30 minutes after the actual treatment point.}
    \label{fig:placebos}
\end{figure}

\paragraph{Placebo Treatment.}
We also implement a placebo treatment analysis. The core logic of this approach is to check whether the estimated treatment effects could be driven by other trends earlier or later than the actual intervention. Shifting the assumed treatment time in 10-minute intervals, ranging from 30 minutes before to 30 minutes after the actual intervention, we re-estimate all DiD models at each placebo point. 

Figure~\ref{fig:placebos} visualises these placebo tests by plotting the precision-weighted average treatment effects and their confidence intervals across all experiments measuring animosity (see Table~\ref{tab:results}). At the actual treatment point $t = 0$, which compares message content in the 30 minutes before and after the first appearance of the trigger word ($[-30,0[$ vs. $[0,30]$), we observe a clear and substantial effect (1.99). We then shift the treatment window rightward in 10-minute steps—first comparing $[-20,10[$ to $[10,40]$, and so on—until the actual intervention fully exits the analysis window at $t = 30$. This generates a steady decline in estimated effects, which converge to zero, as expected if no real treatment occurred at these placebo times. Similarly, shifting the window to earlier times (e.g., comparing $[-60,-30[$ to $[-30,0]$) reveals a decreasing trend in effect size, though some residual difference remains. In conclusion, given the clear peak in the effect at $t=0$, we are confident in the robustness of our approach to defining the actual treatment moment.

\section{Selected Reddit Comments}\label{sec:appendix-example}
Below, we present a Reddit discussion and how it evolves when the trigger ``feminism'' is used. The comment containing the trigger is \textbf{bolded}.


\noindent\textbf{WARNING: the following comments contain sensitive and potentially offensive language. Reader discretion is advised.}

\begin{quote}
"This image though..."
\end{quote}

\begin{quote}
"Haunting for us, isn't it? But not for the females."
\end{quote}

\begin{quote}
"Well I'm sure most females would find this horrifying too."
\end{quote}

\begin{quote}
"You a female? Why are you on this Sub?"
\end{quote}

\begin{quote}
\textbf{"Because I'm against feminism?"}
\end{quote}

\begin{quote}
"Well, it's better if these Posts are seen by all the Men out there. Get the taste of true Chameleon female nature."
\end{quote}

\begin{quote}
"Too late. I already saw it."
\end{quote}

\begin{quote}
"There are many women like us on this sub ."
\end{quote}

\begin{quote}
"Yeah I’m a woman and I don’t like it. Women are not a hive mind."
\end{quote}

\begin{quote}
"Women are not a hive mind. You should go out and talk to one maybe."
\end{quote}

\begin{quote}
"My fucking eyes. Sometimes I wish I was blind."
\end{quote}

\begin{quote}
"When Male dies, it's ""sometimes"" sad; when Female dies.............CRIMEEEEE!!!!
\end{quote}

\begin{quote}
"The fucked up world we live in"
\end{quote}

\end{document}